\documentstyle[12pt]{article}

\newcommand{\be}{\begin{equation}}
\newcommand{\ee}{\end{equation}}
\newcommand{\bd}{\begin{displaymath}}
\newcommand{\ed}{\end{displaymath}}
\newcommand{\bea}{\begin{eqnarray}}
\newcommand{\eea}{\end{eqnarray}}

\newcommand{\half}{\frac{1}{2}}
\newcommand{\lnb}{\mbox{\Large$\left(\right.$}}
\newcommand{\rnb}{\mbox{\Large$\left.\right)$}}
\newcommand{\lsb}{\mbox{\LARGE$\left[\right.$}}
\newcommand{\rsb}{\mbox{\LARGE$\left.\right]$}}

\newcommand{\LCB}{\mbox{\LARGE$\left\{\right.$}}
\newcommand{\RCB}{\mbox{\LARGE$\left.\right\}$}}
\newcommand{\lla}{\mbox{\LARGE$\left\langle\right.$}}
\newcommand{\rra}{\mbox{\LARGE$\left.\right\rangle$}}

\newcommand{\AmS}{{\protect\the\textfont2
  A\kern-.1667em\lower.5ex\hbox{M}\kern-.125emS}}

\title{ Anomalous fermion number non-conservation  \\
        on the lattice }

\author{ Istv\'an Montvay\thanks{Permanent address: DESY, Hamburg, FRG}
\\[1.0em]
                  CERN, Geneva, Switzerland }
\date{}

\begin{document}
\maketitle

\begin{abstract}
The anomaly for the fermion number current is calculated on
the lattice in a simple prototype model with an even number of
fermion doublets.
\end{abstract}


\section{INTRODUCTION}

Fermion number, which is the sum of baryon number and lepton number
($B+L$), is not conserved in the Standard Model \cite{THOOFT}.
This is due to the anomaly in the fermion current.
Under ``normal'' conditions there is, however, a strong suppression
factor
\be \label{eq01}
\exp(-4\pi/\alpha_W) \simeq 10^{-150} \ ,
\ee
which makes $(B+L)$ violation unobservable.
At high temperature and/or high fermion densities (at high energies)
the non-conservation is amplified.
This may explain the small baryon asymmetry of the universe, which
could arise via this mechanism at the cosmological electroweak phase
transition.
(For references to the extensive literature on this subject, see the
reviews in ref. \cite{SHARIN}.)

The lattice formulation of the anomalous fermion number
non-conservation is problematic \cite{BANKS}, because it has to do with
the chiral $\rm SU(2)_L$ gauge coupling and, as is well known,
there is a difficulty with chiral gauge fields on the lattice
(see, for instance, the review \cite{TSUKPR}).
There is, however, an approximation of the electroweak sector of the
Standard Model, which can be studied with standard lattice techniques,
namely the limit when the
$\rm SU(3)_{colour} \otimes U(1)_{hypercharge}$ gauge couplings are
neglected.
The usefulness of this limit for lattice studies was particularly
emphasized in earlier works by Lee and Shrock (see \cite{LEESHR}
and references therein).
In their phase structure studies staggered fermions were used.
Here Wilson fermions will be considered, which naturally lead
to the mirror fermion action for chiral gauge theories \cite{MIRFER}.
There is now a growing amount of experience with this action, without
$\rm SU(2)_L$ gauge field, in the numerical simulation studies of the
allowed region of renormalized quartic and Yukawa couplings
\cite{LINWIT,FLMMPTW,HARLEE}.
The inclusion of the $\rm SU(2)_L$ gauge field in the simulation
algorithms is straightforward; therefore one can start to speculate
about the possibility to explore some features of the violation of the
fermion number conservation.
In order to understand the mechanism of fermion number non-conservation
on the lattice, let us see how the relevant anomalous Ward-Takahashi
identity arises in this formulation.


\section{LATTICE ACTION}

Let us consider a simple prototype model, which is the extension
of the standard $\rm SU(2)_L$ Higgs model by an even number
$2N_f$ of fermion doublets.
In the Standard Model we have $N_f=6$ (for simplicity, we consider
Dirac neutrinos, but the massless neutral right-handed neutrinos
decouple \cite{GOLPET}).
In what follows we take, for simplicity, $N_f=1$, but the extension to
$N_f > 1$ is trivial.
The lattice action depends on the matrix scalar field
$\varphi_x=\phi_{0x}+i\phi_{sx}\tau_s$ (with four real fields
$\phi_{S=0,\ldots,3}$) and the fermion doublet fields $\psi_{(1,2)x}$:
\be \label{eq02}
S=S_{scalar}+S_{fermion} \ .
\ee
The standard Higgs-model action is
\bd
S_{scalar} = \frac{1}{4}\sum_x \LCB
m_0^2 {\rm Tr\,}(\varphi^\dagger_x\varphi_x)
+ \lambda \left[ {\rm Tr\,}(\varphi^\dagger_x\varphi_x) \right]^2
\ed
\be \label{eq03}
+ \sum_{\mu=\pm 1}^{\pm 4}
[ {\rm Tr\,}  (\varphi^\dagger_x \varphi_x)
- {\rm Tr\,}  (\varphi^\dagger_{x+\hat{\mu}}U_{x\mu}\varphi_x) ] \RCB
\ .
\ee
The fermionic part contains the chiral gauge fields
(with $U_{x\mu} \in \rm SU(2)$ and $P_{L,R}=(1 \mp \gamma_5)/2$)
\be \label{eq04}
U_{(L,R)x\mu}=P_{(L,R)} U_{x\mu}+P_{(R,L)}
\ee
and is given by
\bd
S_{fermion} = \sum_x \LCB \frac{\mu_0}{2} \lsb
  (\psi^T_{2x}\epsilon C \psi_{1x}) - (\psi^T_{1x}\epsilon C \psi_{2x})
+ (\overline{\psi}_{2x}\epsilon C \overline{\psi}^T_{1x})
- (\overline{\psi}_{1x}\epsilon C \overline{\psi}^T_{2x}) \rsb
\ed
\bd
-\, \half \sum_\mu \lsb
  (\overline{\psi}_{1 x+\hat{\mu}} \gamma_\mu U_{Lx\mu} \psi_{1x})
+ (\overline{\psi}_{2 x+\hat{\mu}} \gamma_\mu U_{Lx\mu} \psi_{2x})
\ed
\bd
-\, \frac{r}{2} \lnb
  (\psi^T_{2x}\epsilon C \psi_{1x})
- (\psi^T_{2 x+\hat{\mu}}\epsilon C U_{Lx\mu} \psi_{1x})
- (\psi^T_{1x}\epsilon C \psi_{2x})
+ (\psi^T_{1 x+\hat{\mu}}\epsilon C U_{Lx\mu} \psi_{2x})
\ed
\bd
+\, (\overline{\psi}_{2x}\epsilon C \overline{\psi}^T_{1x})
- (\overline{\psi}_{2 x+\hat{\mu}} U_{Rx\mu}
   \epsilon C \overline{\psi}^T_{1x})
- (\overline{\psi}_{1x}\epsilon C \overline{\psi}^T_{2x})
+ (\overline{\psi}_{1 x+\hat{\mu}} U_{Rx\mu}
  \epsilon C \overline{\psi}^T_{2x})
\rnb \rsb
\ed
\be \label{eq05}
+ (\overline{\psi}_{1Rx} G_1\varphi^+_x \psi_{1Lx})
+ (\overline{\psi}_{1Lx} \varphi_x G_1  \psi_{1Rx})
+ (\overline{\psi}_{2Rx} G_2\varphi^+_x \psi_{2Lx})
+ (\overline{\psi}_{2Lx} \varphi_x G_2  \psi_{2Rx}) \RCB \ .
\ee
Here $\epsilon=i\tau_2$ acts in isospin space, and $C$ is the
fermion charge conjugation matrix.
The Yukawa couplings $G_{1,2}$ can, in general, be arbitrary
diagonal matrices in isospin space but, for simplicity, we shall here
only consider the case with degenerate doublets ($G_{1,2}$
proportional to the unit matrix).

Instead of the off-diagonal Majorana mass $\mu_0$ and Majorana-like
Wilson term (proportional to $r$), it is technically more convenient
to consider a Dirac-like form with $\psi \equiv \psi_1$ and the
mirror fermion field
\be \label{eq06}
\chi_x \equiv \epsilon^{-1} C \overline{\psi}_{2x}^T \ ,
\hspace{3em}
\overline{\chi}_x \equiv \psi_{2x}^T \epsilon C  \ .
\ee
In terms of $\psi$ and $\chi$ one obtains the mirror fermion action for
chiral gauge fields \cite{MIRFER}
($G_\psi \equiv G_1,\; G_\chi \equiv G_2$):
\bd
S_{fermion} = \sum_x \LCB
  \mu_0 \left[ (\overline{\chi}_x\psi_x)
+ (\overline{\psi}_x\chi_x) \right]
-\, \half \sum_{\mu = \pm 1}^{\pm 4} \lsb
  (\overline{\psi}_{x+\hat{\mu}} \gamma_\mu U_{Lx\mu} \psi_x)
+ (\overline{\chi}_{x+\hat{\mu}} \gamma_\mu U_{Rx\mu} \chi_x)
\ed
\bd
 - \, r \lnb
  (\overline{\chi}_x\psi_x)
- (\overline{\chi}_{x+\hat{\mu}} U_{Lx\mu}\psi_x)
+ (\overline{\psi}_x\chi_x)
- (\overline{\psi}_{x+\hat{\mu}} U_{Rx\mu} \chi_x) \rnb \rsb
\ed
\be \label{eq07}
+ (\overline{\psi}_{Rx} G_\psi\varphi^\dagger_x \psi_{Lx})
+ (\overline{\psi}_{Lx} \varphi_x G_\psi  \psi_{Rx})
+ (\overline{\chi}_{Lx} G_\chi\varphi^\dagger_x \chi_{Rx})
+ (\overline{\chi}_{Rx} \varphi_x G_\chi  \chi_{Lx}) \RCB \ .
\ee
This is the appropriate form of the fermion action in the phase with
broken symmetry, as the investigations of the corresponding chiral
Yukawa models show \cite{FKLMMM,LMMW,LINWIT,FLMMPTW,HARLEE}.

In the symmetric (i.e.~confinement) phase, however, there is
a natural alternative choice in terms of the reshuffled combinations
\cite{FKLMMM}:
\be \label{eq08}
\psi_{Ax} \equiv \psi_{Lx}+\chi_{Rx} \ , \hspace{2em}
\psi_{Bx} \equiv \chi_{Lx}+\psi_{Rx} \ .
\ee
On this basis the vector-like nature of the model becomes explicit
($\gamma_5$'s appear only in the Yukawa couplings).
The $\rm SU(2)$ gauge field couples only to $\psi_A$, and the
neutral doublet $\psi_B$ has only its Yukawa coupling.

Note the different r\^oles played by $\mu_0$ in the three lattice
actions: in (\ref{eq05}) it is an off-diagonal Majorana mass, in
(\ref{eq07}) the fermion-mirror-fermion mixing mass, whereas on the
basis in (\ref{eq08}) it is a common Dirac mass for $\psi_A$ and
$\psi_B$.
The physical interpretation of the model is, of course, given in terms
of the action in (\ref{eq05}).

Previous studies of the phase structure of the same continuum ``target
theory'' in the staggered fermion formulation were usually done in a
basis corresponding to (\ref{eq08}), with the known differences between
staggered and Wilson fermions (see, for instance,
\cite{LEESHI,AOLESH,LEESHR,KUTI}).
In many cases the Yukawa couplings were omitted, and the $\psi_B$
field was not considered at all.
Representing the fermion number anomaly both in terms of the fields in
(\ref{eq05}) and (\ref{eq08}) is useful, because it gives the
connection to the axial anomaly.
This connection has recently been exploited also in ref. \cite{MAGSHI}.

\section{THE ANOMALY}

On smooth background scalar and gauge fields $\{\varphi_x,U_{x\mu}\}$
the effective action is defined by
\be \label{eq09}
\exp\{ -S_{eff}[U,\varphi] \} \equiv
Z_{f0}^{-1} \int [d\overline{\Psi} d\Psi]
\exp \{-S_f[\Psi,\overline{\Psi},U,\varphi] \} \ ,
\ee
where $\Psi \equiv \{ \psi,\chi \}$.
An infinitesimal fermion number transformation is:
\bd
\psi_x = (1+i\alpha_x)\psi_x^\prime \ ,  \hspace{2em}
\chi_x = (1-i\alpha_x)\chi_x^\prime \ ,
\ed
\be \label{eq10}
\overline{\psi}_x = (1-i\alpha_x)\overline{\psi}_x^\prime \ ,
\hspace{2em}
\overline{\chi}_x = (1+i\alpha_x)\overline{\chi}_x^\prime \ .
\ee
This corresponds to the fact that the fermion number is defined to be
$+1$ for the fields $\psi_{1,2}$ (and hence it is $-1$ for $\chi$).

The gauge-invariant fermion number current can be defined as
\be \label{eq11}
J_{x\mu} \equiv \frac{1}{2} \lsb
(\overline{\psi}_{x+\hat{\mu}} \gamma_\mu U_{Lx\mu} \psi_x)
+(\overline{\psi}_x \gamma_\mu U_{Lx\mu}^\dagger \psi_{x+\hat{\mu}})
-(\overline{\chi}_{x+\hat{\mu}} \gamma_\mu U_{Rx\mu} \chi_x)
-(\overline{\chi}_x \gamma_\mu U_{Rx\mu}^\dagger \chi_{x+\hat{\mu}})
\rsb \ .
\ee
Introducing the new integration variables
($\psi^\prime$, $\overline{\psi}^\prime$,
 $\chi^\prime$, $\overline{\chi}^\prime$)
in the path integral with action (\ref{eq07}), one obtains with
$\Delta^b_\mu f_x \equiv f_x - f_{x-\hat{\mu}}$
the lattice W-T identity
\bd
\langle \Delta^b_\mu J_{x\mu} \rangle_f = \lla
2\mu_0 [(\overline{\chi}_x\psi_x) - (\overline{\psi}_x\chi_x)]
\ed
\bd
+\, \frac{r}{2} \sum_{\mu=1}^4
\lsb 4(\overline{\chi}_x\psi_x)
-(\overline{\chi}_{x+\hat{\mu}} U_{Lx\mu} \psi_x)
-(\overline{\chi}_x U_{Lx\mu}^\dagger \psi_{x+\hat{\mu}})
\ed
\bd
-\, (\overline{\chi}_x U_{Lx-\hat{\mu},\mu} \psi_{x-\hat{\mu}})
-(\overline{\chi}_{x-\hat{\mu}} U_{Lx-\hat{\mu},\mu}^\dagger \psi_x)
- 4(\overline{\psi}_x\chi_x)
+(\overline{\psi}_{x+\hat{\mu}} U_{Rx\mu} \chi_x)
+(\overline{\psi}_x U_{Rx\mu}^\dagger \chi_{x+\hat{\mu}})
\ed
\be \label{eq12}
+(\overline{\psi}_x U_{Rx-\hat{\mu},\mu} \chi_{x-\hat{\mu}})
+(\overline{\psi}_{x-\hat{\mu}} U_{Rx-\hat{\mu},\mu}^\dagger \chi_x)
\rsb \rra_f \ .
\ee
This has to be evaluated in the continuum limit, when the
momenta of the external fields in lattice units are of the order
$a\;\; (a \to 0)$.

For small lattice spacing $a$ the left-hand side of (\ref{eq12})
is of order $a^4$ (note that for the moment we keep the bare
parameters fixed, for instance, $\mu_0$ can be of order 1). Therefore
diagrammatically the contributing graphs can have at most four
external field legs.
Explicit evaluation shows that in the present case only those with
two or three external fields (i.e.~the triangle and quadrangle graphs)
contribute.
Introducing the $\rm SU(2)$ field strength as usual by
\be \label{eq13}
F_{\mu\nu}^s(x) = \partial_\mu A_\nu^s(x) - \partial_\nu A_\mu^s(x)
+ g\epsilon_{stu} A_\mu^t(x) A_\nu^u(x) \ ,
\ee
the result is (this time for $2N_f$ fermion doublets)
\be \label{eq14}
\langle \partial_\mu J_\mu(x) \rangle_f =
\lim_{a \to 0} \langle \Delta^b_\mu J_{x\mu} \rangle_f a^{-4}
= N_f g^2\epsilon_{\mu\nu\rho\sigma}
F_{\mu\nu}^s(x)F_{\rho\sigma}^s(x) {\cal I}(r,\mu_0) \ .
\ee
Here the lattice integral ${\cal I}$ is given by
\be \label{eq15}
{\cal I}(r,\mu_0) \equiv \frac{1}{(2\pi)^4} \int_{-\pi}^\pi
\frac{\mu_k \cos k_1 \cos k_2 \cos k_3 \cos k_4}
{(\bar{k}^2 + \mu_k^2)^3}
\cdot
\lsb r \sum_{\alpha=1}^4
\bar{k}_\alpha^2/\cos k_\alpha - \mu_k \rsb d^4 k\ ,
\ee
and the notations are
\bd
\mu_k = \mu_0 + \frac{r}{2} \hat{k}^2 \ ,
\ed
\be \label{eq16}
\bar{k}_\mu = \sin k_\mu \ , \hspace{1em}
\hat{k}_\mu = 2\sin \frac{k_\mu}{2} \ .
\ee

The integral ${\cal I}$ is the same as the one occurring in the chiral
anomaly, and one can prove (see e.g.~\cite{KARSMI,SEISTA})
\be \label{eq17}
{\cal I}(r,0) = \frac{1}{32\pi^2} \hspace{2em}
({\rm independently\; from}\; r) \ .
\ee
Note that in the present regularization scheme no other terms on the
right-hand side of (\ref{eq14}) occur.
For instance, the scalar field having Yukawa couplings to the
fermions does not contribute at all (although, of course, it appears
on external legs of the graphs).
This is different from the non-Abelian $\rm U(N) \otimes U(N)$
anomaly studied in ref. \cite{COKONA} in other regularization
schemes (with different lattice actions), where the Bardeen-counterterms
\cite{BARDEE} are in general non-zero.

Equations (\ref{eq14}) and (\ref{eq17}) show that the correct continuum
anomaly is reproduced at vanishing bare (Majorana) fermion mass
$\mu_0=0$.
It is, however, interesting to investigate the $\mu_0$ dependence
of the lattice integral in (\ref{eq15}).
The numerical evaluation of the corresponding lattice sum
${\cal I}_L$ on $L^4$ lattices up to $L=200$ shows that
${\cal I} = \lim_{L \to \infty} {\cal I}_L$ is very small,
probably ${\cal I}(\mu_0,r)=0$ for every positive $\mu_0$ \cite{FERMAN}.
This behaviour implies that the anomaly in (\ref{eq14})
disappears at every positive $\mu_0$, and there is a singularity at
$\mu_0=0$, where according to (\ref{eq17}) the value of ${\cal I}$ is
non-zero.
The derivatives of ${\cal I}$ with respect to $\mu_0$ tend to
infinity for $L \to \infty$; therefore we have, on the given external
bosonic field configuration, for instance,
\bd
-\frac{\partial}{\partial\mu_0} \langle \Delta^b_\mu J_{x\mu} \rangle_f
= \langle \Delta^b_\mu J_{x\mu} \sum_y [(\overline{\chi}_y \psi_y)
+(\overline{\psi}_y \chi_y)] \rangle_f
\ed
\be \label{eq18}
\hspace{5cm}
- \langle \Delta^b_\mu J_{x\mu} \rangle_f
\lla \sum_y [(\overline{\chi}_y \psi_y)
+(\overline{\psi}_y \chi_y)] \rra_f \to \infty \ .
\ee
Here the infinity can be produced by the summation over $y$
because of the long-range correlation due to fermionic zero modes.

{}From the practical point of view the behaviour of ${\cal I}(r,\mu_0)$
implies that in numerical simulations one has to be careful in the
extrapolation to $\mu_0=0$.
The lattice volume should be small enough.

The functional dependence of the lattice integral ${\cal I}(r,\mu_0)$
on $\mu_0$ also illustrates how the anomaly is emerging from the
explicit symmetry breaking present in the lattice action (\ref{eq02}).
In the case when the cut-off can be completely removed, this does not
matter.
Nevertheless, in theories with scalar fields there is the well-known
``triviality problem'', which implies that for finite renormalized
couplings the lattice spacing cannot be taken to zero.
This means that the unpleasant feature of the singular dependence of the
anomaly on the bare fermion mass in principle remains.
However, in order to understand the situation better, one has to
consider the full theory with quantized bosonic fields, where
a mixing of the renormalized composite operators has to be dealt
with \cite{ESPTAR,SMIVIN}.

{\bf Acknowledgements:} It is a pleasure to thank L.~Alvarez-Gaum\'e,
A.~Ringwald and M.~Sha\-posh\-ni\-kov for enlightening discussions, and
the Theory Division of CERN for hospitality during the preparation of
this proceedings contribution.


\end{document}